# The Effect of Funding on Student Achievement: Evidence from District of Columbia, Virginia, and Maryland[1]

Adam Raabe[1], Jessica Reynolds[1], Akshitha Kukudala[1], and Huthaifa Ashqar[1,2]

[1]Department of Data Science, University of Maryland Baltimore County

[2]Arab American University, huthaifa.ashqar@aaup.edu

**Abstract**

The question of how to best serve the student populations of our country is a complex topic. Since public funding is limited, we must explore the best ways to direct the money to improve student outcomes. Previous research has suggested that socio-economic status is the best predictor of student achievement, while other studies suggest that the amount of money spent on the student is a more significant factor. In this paper, we explore this question and its impacts on Maryland, Virginia, and the District of Columbia schools. We conclude that the graduation rate has a direct relationship with unemployment, suggesting that funding towards improving out-of-school opportunities and quality of life will significantly improve students' chances of success. We do not find a significant relationship between per-pupil spending and student achievement.

**Introduction**

The relationship between school funding and student achievement is hotly debated among scholars. Some contend that the effect of funding on student achievement is negligible in comparison to other factors (Coleman, 1966). Others claim to have found a link between funding and increased student achievement (Jackson et al., 2015), while still others claim that funding is only helpful when it is designated to certain aspects of the school system (Wenglinsky, 1997). While research supports all arguments, one question that could use more investigation is that of opportunity: Does a successful student have an opportunity to look forward to a meaningful and successful life after completing school? What role does such opportunity play in motivating students to achieve a higher standard? In this study, we explore this question, to challenge the idea that funding allocation to schools is the most significant usage of that money.

---

[1] All resources and code used for this study can be found at https://github.com/araabe2/Data-601-Group-Project.

A great deal of effort has been put into understanding the complex set of circumstances that influences student achievement. Various studies have attempted to establish a consensus on both sides of the discussion, with varying degrees of success. The discussion itself is generally agreed to have begun with the 1966 Coleman study that determined that, in large part, the greatest influence on student achievement is that of familial socio-economic status (Coleman et al., 1966). This raised questions about the true nature of funding's effect on student achievement and has subsequently been investigated numerous times. Many articles determined a positive relationship between money funding directed to schools and student achievement, such as one that determines that an additional $1,000 of spending per student could reduce dropout rates by 2% (Kreisman & Steinberg, 2019). Others found the opposite results, such as research done in Michigan schools on a variety of factors, concluding that there was no significant relationship between student achievement and per-pupil spending (Snyder, 1995). Others, still, posited that students who received the benefit of an additional 10% of funding for all 12 years of schooling would have significant increases in quality-of-life indicators after completing school, an effect they claim to be heightened in children from low-income families (Jackson et al., 2015). Yet others found that funding benefits were short-lived, finding that additional funding to schools in Massachusetts as the result of educational reform caused improvement in 4$^{th}$-graders' test scores, but was no longer showing any effect on 8$^{th}$-graders' test scores (Guryan, 2001). The research has been staunchly obscure, despite researchers' best efforts.

Several researchers have taken a different approach to the question: What if the reason that the research has been unable to clarify the issue is that the variation is caused by the way in which the money is spent? Research by Wenglinsky indicates that not all spending has equivalent effects, and that the lack of consideration of this has been confounding the discussion up to that point (Wenglinsky, 1997). Research into meta-analysis of the literature developed on the topic seems to support this, concluding that there is an average increase in student achievement due to increased spending, but not in all cases, notably not in the case of Title I spending (Jackson, 2018). Other research focuses on the impact of other predictors on student achievement, such as school environment or class size, which are, in turn, affected by choices in educational funding allocation. Continuing research into the Coleman data using modern methods found that as much as 40% of the difference in student achievement was due to differences in school environment, while admitting that the greatest amount of variability is still within-school (Konstantopoulos & Borman, 2011). Research investigating the effect of class sizes on achievement gaps between Caucasian and African-American students determined that even short-lived periods of decreased class size lead to lasting growth in student achievement, more pronounced in the African-American students (Krueger & Whitmore, 2001), research supported by the Wenglinsky study (1997). Research into the effect that school culture has on student achievement in Texas schools found that student achievement is a strong indicator for healthy school culture, defined on metrics such as morale, communication, and cohesiveness (MacNeil et al., 2009). Studies such as these suggest that intentional effort should be put into considering how best to allocate funding for maximum benefit to the students.

In this study, we explore the relationships between several variables: student achievement, general quality of life, and funding that influences students. To represent these factors, we considered several alternatives for each, eventually settling on 4-year graduation rate, unemployment, and per-pupil total spending, respectively. Our hypotheses are twofold:

a) There is a relationship between unemployment and graduation rate.
b) The relationship between unemployment and graduation rate is a stronger predictor of graduation rate than the relationship between per-pupil spending and graduation rate.

We used data from the states of Virginia and Maryland, as well as the District of Columbia. In these regions, school districts and counties coincide, except in special cases. These regions also represent a diverse collection of district types, ranging from cities to rural towns and from places with dense populations to low populations. The states as a whole also represent a wide spread of the unemployment range, with Virginia being ranked $15^{th}$ in the US for least unemployment, Maryland ranking $42^{nd}$, and DC ranking $51^{st}$, as of March 2022 (US Bureau of Labor Statistics).

**Dataset**

The data used in our analysis was agglomerated from several locations. We chose Maryland, Washington DC, and Virginia to be our data scope due to the overlap of school district and county, which allowed us to match public data about individual counties to the school districts they also represent. We primarily used three sources, one for each type of data considered. Unemployment data was sourced from the public data on the US Bureau of Labor Statistics. We took data for each year from 2015 to 2021, although we ultimately only used data from 2015 to 2018, because coincidental data for other analysis categories could not be retrieved. While the data on this site is granular to the level of month, for the sake of consistency we chose December's unemployment numbers to represent each year.

Data used to represent student achievement was sourced from County Health Ranks & Roadmaps, a website that, in turn, sources their information from EDFacts, information publicly available from the US Department of Education. This data represents the number of students who graduated after 4 years of highschool with the correct anticipated graduation cohort determined by their age. It includes information both about cohort size and graduation rate, broken down by year. Due to the nature of this data, students who graduated, but may have taken an extra year to complete highschool are not considered at any point.

Data used to identify per-pupil spending was sourced from the website for the US Census Bureau. This website houses data about school districts, reported federally by the states, who receive their information from the schools. The data itself contains a vast range of characteristics, such as federal funding received, student population, total funding spent on teacher salaries, and type of school locale served (town, large city, small city, rural, etc.). Ultimately, we decided that the most helpful characteristic that we would consider was that of "Total Per Pupil Spending," which should be equivalent to total revenue divided by the number of enrolled students. Other characteristics were added to the data to facilitate further research.

**Methods**

To audit the relative strength of each variable as a predictor, we decided to test the linear relationship between them. Linearity between variables would imply that there is a relationship that ties them together, and following that, we can look at the coefficient of determination to understand how well the model fits the data and the p-values to evaluate if the coefficient is statistically significant from zero. A lack of linearity would imply that the variables are unrelated or are better represented by a different model. To determine whether this was a plausible possibility, we initially began by analyzing the data through the use of scatterplots, attempting to assess patterns. Through this process, we determined that there was no obvious alternative relationship between any of the predictors.

We output several different linear analyses. Our first model analyzed the relationship between unemployment and graduation rates using data from 148 counties across 4 years, our second the relationship between per-pupil spending and graduation rates using data from 149 counties across 4 years. In continuing exploration efforts to add richness to the explanation of the results that we were getting, we also conducted a linear regression for the estimated percent of the community with three or more risk factors and graduation rates using data from 147 counties in 2019.

**Analysis and Results**

Our initial investigation into the relationships between our variables took the form of visual analysis of the graph of graduation rate and per-pupil spending (see Figure 1) and that of graduation rate and unemployment (see Figure 2). We noted a few data points in Figure 1 that seemed to be outliers, linking them to Worcester County, Maryland. Further research into the unemployment data for the county revealed that the unemployment rate fluctuated an average of 8.25 percentage points between the summer and winter months, a much higher range than the other counties in the scope of our research (US Bureau of Labor Statistics, 2022). Since the decision to use December for our unemployment data was not representative of Worcester County throughout the year, we decided to drop the data in our analysis of the impact of unemployment rates.

First, we compared graduation rates to per pupil spending and found that the conditions of homoscedasticity and normality needed for a linear regression may be violated. Not surprisingly, the estimated coefficient for per pupil spending was not found to be statistically significant and the coefficient of determination is very weak ($R^2 = 0.0034$, $F(1, 594)=2.018$, $p=0.156$). Next, we compared graduation rates to unemployment rates. Here all assumptions needed for linear regression appeared to be met. The estimated coefficient showed that an increase of 1 percentage point in unemployment results in a 2 percentage point drop in graduation rate and was statistically significant ($R^2 = 0.126$, $F(1, 590)=84.95$, $p<0.001$). Finally, we compared graduation rates with the percent of the population with 3 or more risk factors. Again, the tests for the assumptions needed for linear regression did not show any violations. Due to the spread of the date, the coefficient of determination is not as strong, but showed that there is a statistically significant coefficient where for each percentage point increase in the population with three or more risk factors, there is a predicted drop in graduation rate of 0.25 percentage points ($R^2 = 0.0875$, $F(1, 146)=14.01$, $p<0.001$).

**Figure 1**

*Graduation Rate vs Per Pupil Spending by General County Type*

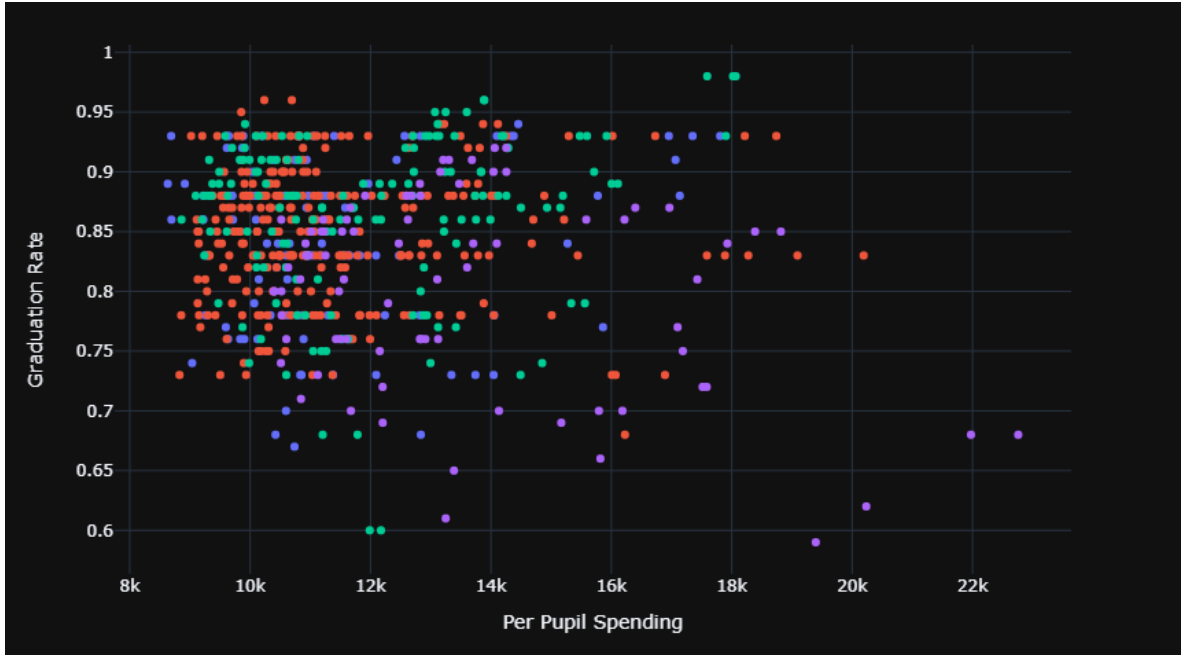

**Figure 2**

*Graduation Rate vs Unemployment Rate by General County Type*

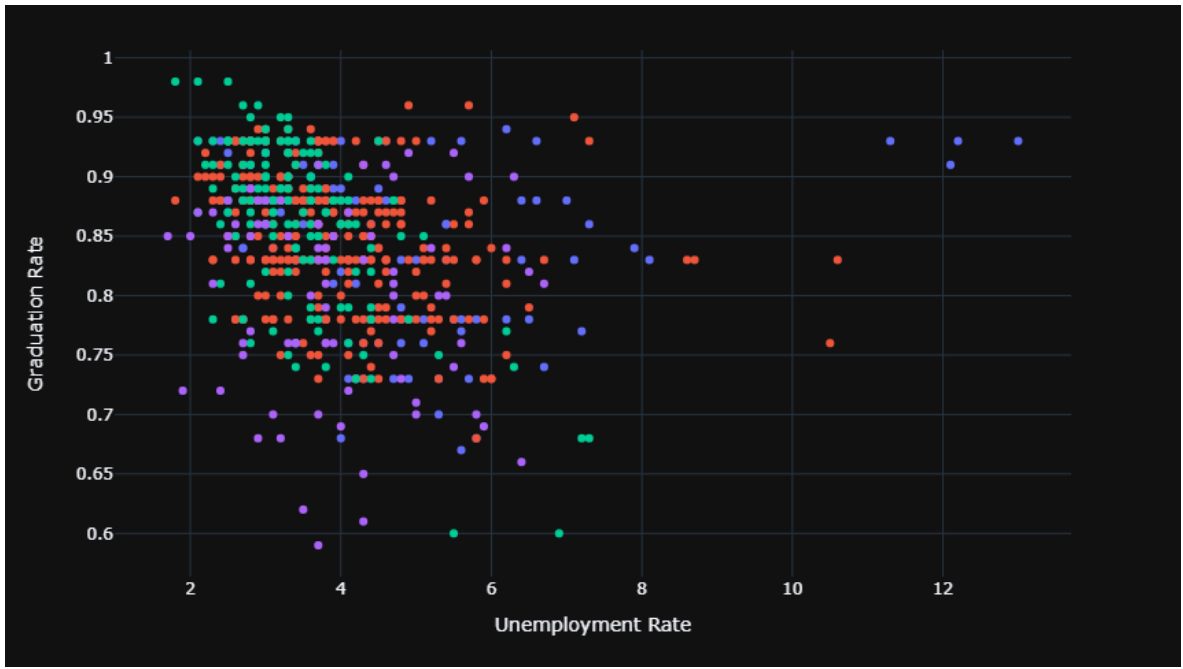

*Note*. The colors of the scatter plots represent different general county types. Green = suburban, blue = town, red = rural, and purple = city.

**Discussion**

Within our analysis, we noticed several anomalies that we were unable to address. First, we noticed that the data gathered for graduation rates for 2017 and 2018 are identical, which could impact the reliability of our models. Second, while our decision to focus on just Maryland, Virginia, and the District of Columbia did represent every type of county, this resulted in a large cluster of data in the $8,000 to $12,000 range for per pupil spending, when our source data seems to indicate that there are school districts elsewhere in the country with as low as $340 per student. Third, the costs associated with running a school are also dependent on the location, for which we were not able to account within the data we collected. Fourth, while our models can shed light on factors that are more highly correlated with graduation rates, due to the spread of the data they should not be used for making predictions. Lastly, our analysis shows that there seems to be a slight negative correlation between money spent on each student and how likely they are to graduate, but this does not take into account that more time may be needed before the benefits of the increase in investment can be measured, nor is it statistically significant.

We performed an exploratory analysis to attempt to explain why we might have seen a negative trend. We considered the possibility that perhaps schools that are performing lower might receive extra funding in an attempt to bolster their efforts to improve. If this were the case, we would expect to see higher growth in schools with higher funding. Adam's investigation seemed to suggest that this is not the case. Schools that received an average funding of less than $14,000 per pupil had an average growth rate of 1.4%, while schools that received an average funding of more than $14,000 per pupil only had an average growth rate of 0.75%. Still, we think a continued exploration upon this facet might provide different results.

While we chose specific representative statistics for predicting graduation rates, there are others that hold promising potential as well. Considering alternative indicators for academic success, like SAT or AP scores, might be relevant due to their relationship to how students will perform in college. Rates of failure or other quantifiers of average scores in benchmark classes (like algebra or reading) likely correlate to graduation rates. Also, considering alternative public quality-of-life indicators, such as median or average income, cost of living, crime rate, college attendance, single parent households, and income inequality, might lead to even stronger predictors for student success.

**Conclusion**

Our research shows a significant and negative correlation between unemployment and graduation rate, with each additional percent of unemployment predicting a drop-in graduation rate by nearly 2%. Since we were unable to establish a significant relationship between graduation rate and per-pupil spending, we determine that both of our hypotheses are confirmed. It is critical to address the current inequities in academic achievement among students but we have an obligation to taxpayers to ensure that public funding is being used efficiently to address the issues that have the greatest impact. Our analysis shows that graduation and unemployment rates can be modeled as a linear relationship and have a much higher correlation than per pupil spending. This aligns with the findings of the 1966 Coleman study that pointed to familial and socio-economic status to have the greatest influence on student success in school (Coleman et al., 1966). Thus, investment in programs that address unemployment and other quality of life factors may have a greater impact on improving educational outcomes for students than increased spending on schools. Given that graduation rates and percentage of the population with three or more risk factors are also more highly correlated than per pupil spending, further analysis of the impact of income-to-poverty ratio, living conditions, family academic achievement, access to internet, and communication barriers on student outcomes will shed light on where to focus future public investment.

# References

Ashqar, H., Chon, C., Gilmer, L., McMillan, A., Nagorniuk, A., Walbridge, T., O'Connor, R., Woodson, C., Lyon-Hill, S., Provo, J. and Tate, S., 2022. Regional Economic Recovery and Resilience Toolkit.

Coleman, J. S., et al. (1965, November 30). *Equality of educational opportunity.* ERIC. Retrieved May 22, 2022, from https://eric.ed.gov/?id=ED012275

Guryan, J. (2001). Does money matter? Regression-discontinuity estimates from education finance reform in Massachusetts. *Working Paper 8269*. https://doi.org/10.3386/w8269

Jackson, C. K. (2018). Does school spending matter? The new literature on an old question. *Working Paper 25368*. https://doi.org/10.3386/w25368

Jackson, C. K., Johnson, R., & Persico, C. (2015). The effects of school spending on educational and economic outcomes: Evidence from school finance reforms. *Working Paper 20847*. https://doi.org/10.3386/w20847

Konstantopoulos, S., & Borman, G. D. (2011). Family background and school effects on student achievement: A multilevel analysis of the Coleman data. *Teachers College Record: The Voice of Scholarship in Education*, *113*(1), 97–132. https://doi.org/10.1177/016146811111300101

Kreisman, D., & Steinberg, M. P. (2019). The effect of increased funding on student achievement: Evidence from Texas's Small District Adjustment. *Journal of Public Economics*, *176*, 118–141. https://doi.org/10.1016/j.jpubeco.2019.04.003

Krueger, A. B., & Whitmore, D. M. (2001). Would Smaller Classes Help Close the Black-White Achievement Gap? *Princeton Working Paper #451*. Retrieved May 22, 2022, from https://www.semanticscholar.org/paper/Would-Smaller-Classes-Help-Close-the-Black-White-Krueger-Whitmore/96bca843b9dbec29897682e87b8e777e637ec882#paper-header.

MacNeil, A. J., Prater, D. L., & Busch, S. (2009). The effects of school culture and climate on student achievement. *International Journal of Leadership in Education*, *12*(1), 73–84. https://doi.org/10.1080/13603120701576241

Masri, S., Raddad, Y., Khandaqji, F., Ashqar, H. I., & Elhenawy, M. (2024). Transformer Models in Education: Summarizing Science Textbooks with AraBART, MT5, AraT5, and mBART. arXiv preprint arXiv:2406.07692.

Snyder, D. H. (1995, April). *A study of the correlation between per pupil funding and student achievement in the State of Michigan*. ScholarWorks at WMU. Retrieved May 22, 2022, from https://scholarworks.wmich.edu/dissertations/1773/

U.S. Bureau of Labor Statistics. (2022, May 20). 2022 Unemployment Rates for States. U.S. Bureau of Labor Statistics. Retrieved May 22, 2022 from https://www.bls.gov/web/laus/laumstrk.htm

*Unemployment rate in Worcester County, MD*. FRED. (2022, April 27). Retrieved May 22, 2022, from https://fred.stlouisfed.org/series/MDWORC7URN

Wenglinsky, H. (1997). How money matters: The effect of school district spending on academic achievement. *Sociology of Education*, *70*(3), 221. https://doi.org/10.2307/2673210

Whieldon, Lee, and Huthaifa I. Ashqar. "Predicting residential property value: a comparison of multiple regression techniques." SN Business & Economics 2, no. 11 (2022): 178.

**Data sources**

County Health Rankings & Roadmaps. (2022, March). Retrieved March 8, 2022, from
https://www.countyhealthrankings.org/app/

U.S. Bureau of Labor Statistics. (n.d.). *Local Area Unemployment Statistics Map*. U.S. Bureau of Labor Statistics. Retrieved March 8, 2022, from
https://data.bls.gov/lausmap/showMap.jsp;jsessionid=98AD291A95B4F55B407C733AA00D67E2._t3_08v

United States Census Bureau. (2021, August). *Community Resilience Estimates* . Census.gov. Retrieved March 8, 2022, from
https://data.census.gov/cedsci/table?g=0400000US51%240500000&d=CRE%2BCommunity%2BResilience%2BEstimates

United States Census Bureau. (2021, October 8). *Annual survey of school system finances tables*. Census.gov. Retrieved March 8, 2022, from https://www.census.gov/programs-surveys/school-finances/data/tables.html